\documentclass[twocolumn,aps,prb,floatfix,psfig,superscriptaddress]{revtex4-1} 
\usepackage{amssymb,amsmath} 
\usepackage{color}
\usepackage{multirow}
\usepackage{braket}
\usepackage{tabularx,booktabs}  
\usepackage{graphicx}
\usepackage{dcolumn}
\usepackage[section]{placeins}
\usepackage{bm}
\usepackage{upgreek}
\usepackage{marvosym }
\usepackage{mathtools}
\usepackage{float}
\usepackage{latexsym,epsfig}  
\newcommand{\be}{\begin{equation}}
\newcommand{\ee}{\end{equation}} 
\newcommand{\bea}{\begin{eqnarray}}   
\newcommand{\eea}{\end{eqnarray}}

\begin{document}
\title{
Emergence of giant orbital Hall and tunable spin Hall effects in centrosymmetric TMDs\\
}
\author{Pratik Sahu}
\email{pratiksahu2413@gmail.com}
\affiliation{Center for Atomistic Modelling and Materials Design, Indian Institute of Technology Madras, Chennai 600036, India } 
\affiliation{Condensed Matter Theory and Computational Lab, Department of Physics, Indian Institute of Technology Madras, Chennai 600036, India }

\author{Jatin Kumar Bidika}
\affiliation{Center for Atomistic Modelling and Materials Design, Indian Institute of Technology Madras, Chennai 600036, India } 
\affiliation{Condensed Matter Theory and Computational Lab, Department of Physics, Indian Institute of Technology Madras, Chennai 600036, India }

\author{Bubunu Biswal}
\affiliation{Center for Atomistic Modelling and Materials Design, Indian Institute of Technology Madras, Chennai 600036, India } 
\affiliation{Condensed Matter Theory and Computational Lab, Department of Physics, Indian Institute of Technology Madras, Chennai 600036, India }

\author{S. Satpathy}
\email{satpathys@missouri.edu}
\affiliation{Department of Physics \& Astronomy, University of Missouri, Columbia, MO 65211, USA} 
\affiliation{Condensed Matter Theory and Computational Lab, Department of Physics, Indian Institute of Technology Madras, Chennai 600036, India }

\author{B. R. K. Nanda}
\email{nandab@iitm.ac.in}
\affiliation{Center for Atomistic Modelling and Materials Design, Indian Institute of Technology Madras, Chennai 600036, India }
\affiliation{Condensed Matter Theory and Computational Lab, Department of Physics, Indian Institute of Technology Madras, Chennai 600036, India }

\begin{abstract}
We demonstrate the formation of orbital and spin Hall effects (OHE/SHE) in the 1T phase of non-magnetic transition metal dichalcogenides. With the aid of density functional theory calculations and model Hamiltonian studies on MX$_2$ (M = Pt, Pd and X = S, Se, and Te), we show an intrinsic orbital Hall conductivity  ($\sim 10^3 \hbar /e\ \Omega^{-1}cm^{-1}$) , which primarily emerges due to the orbital texture around the valleys in the momentum space. The robust spin-orbit coupling in these systems induces a sizable SHE out of OHE. Furthermore, to resemble the typical experimental setups, where the magnetic overlayers produce a proximity magnetic field, we examine the effect of magnetic field on OHE and SHE and showed that the latter can be doubled in these class of compounds.  With a giant OHE and tunable SHE, the 1T-TMDs are promising candidates for spin and orbital driven quantum devices such as SOT-MRAM, spin nano-oscillators, spin logic devices etc., and to carry out spin-charge conversion experiments for fundamental research.



\end{abstract}   
\date{\today}			 		
\maketitle
\section{Introduction}

The generation of transverse orbital/spin/charge current with a longitudinal applied electric field is known as the orbital/spin/anomalous Hall effect (OHE/SHE/AHE)\cite{she-1,she-2,she-3}. These transverse quantum phenomena have drawn significant attention due to their potential application in designing futuristic quantum devices. While the AHE can be observed by simply measuring the transverse charge current, in the case of OHE and SHE, the transverse magnetic moments are generally measured by passing orbital/spin current through a ferromagnet. The OHE is a fundamental effect, in the sense that it arises from the intrinsic orbital moment and does not require external magnetic field or interaction. On the other hand, the SHE is either intrinsically generated in a magnetic material or in a spin orbit coupling (SOC) active non magnetic material through SOC. In the latter case, the OHE induces a spin current via SOC\cite{Jo,Go,sayantika_1,sayantika_2,pratik-ohe}. Interestingly, several theoretical works\cite{Jo,Go,sayantika_1,sayantika_2,pratik-ohe} have proposed that the OHE is very large and is the dominant one in some materials. 
\begin{figure}[!htb]
    \centering
    \includegraphics[width=0.265\textwidth]{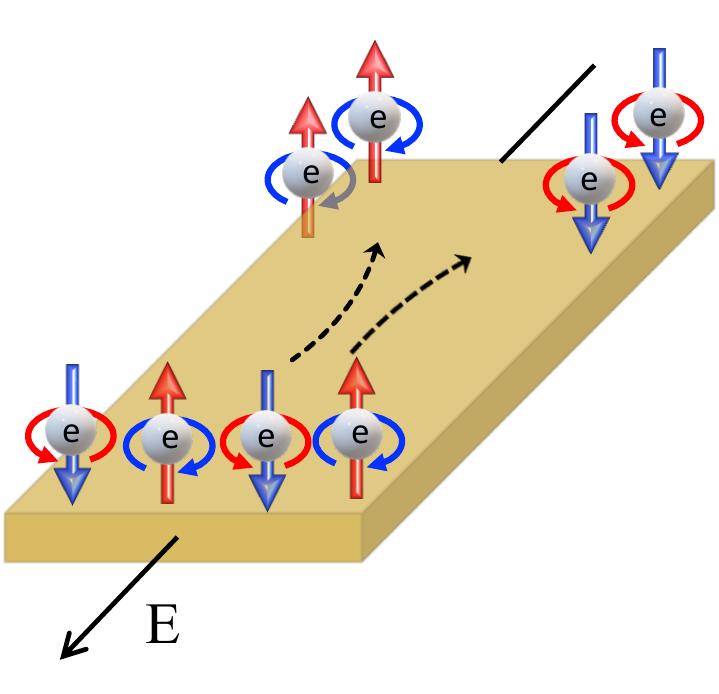}
    \caption{A schematic diagram illustrating orbital and spin Hall effect, where the electric field is applied along the x-axis, and the orbital/spin current is produced along the y-axis. Vertical arrows represent spins, while the circles indicate orbital magnetization.} 
    \label{fig:enter-label}
\end{figure}

In most of the experiments, OHE and SHE are manifested through the spin orbit torque (SOT), which arises due to the spin current\cite{ot-1,ot-2,ot-3}. Therefore, in the context of non-magnetic materials, it is prudent to explore materials with large OHE and a strong SOC. For the former, the non centrosymmetric systems are natural choices as the absence of inversion symmetry allows the spatial accumulation of orbital moments. From the materials point of view, keeping the large SOC in mind, Pt has been reported to exhibit large OHE and SHE. It has also been found that several low SOC metals show huge OHE even if the SHE is negligible. Such materials are of great interest as the large orbital current gives rise to strong SOT. In this context, the non-centrosymmetric monolayer TMDs are promising \cite{sayantika_1,sayantika_2,rap1,rap2}. However, it is of practical interest to explore the  bulk TMDs as potential candidates for SOT. However, they are centrosymmetric and as a consequence, the quenching of the intrinsic orbital moment occurs leading to negligible OHE.  

In this work, we carried out both density functional and model Hamiltonian calculation to propose that the bulk centrosymmetric TMDs (1T-MX$_2$; M = Pt and Pd, X = S, Se, and Te) can exhibit large OHE, when they are subjected to an applied electric field. In addition, we also consider an external magnetic field term in our model Hamiltonian to study the OHE, SHE and AHE. This resembles the proximity magnetic field produced by the ferromagnetic overlayers, which are usually used in the experimental setups\cite{OHE-3-Cr, OHE-4-Ti, pranaba}. 

Our findings suggest that the OHE ($\simeq 800-1600 \ \hbar/e\ \Omega^{-1} cm^{-1}$) is nearly two order higher than the SHE ($\simeq 0.1-38\ \hbar/e\ \Omega^{-1} cm^{-1}$) for the 1T-MX$_2$. However, as the magnetic field is turned on, the SHE is nearly doubled, while the OHE changed a little. This result is significant since the spin current directly gives rise to SOT. Furthermore, since the OHE to SHE conversion factor, while the orbital current passes through a ferromagnetic overlayer\cite{ot-1}, is 
$\simeq 2.5\%$, the large OHE itself provides a sizable contribution to the SOT. Hence, the present study opens up possibilities to tune the SOT experimentally with the aid of ferromagnetic overlayer, which provides a proximity magnetic field to the non-magnetic material\cite{pranaba}.

\section{Basic formalism and computational method}

\begin{figure}[!htb]
\includegraphics[width = 0.5\textwidth]{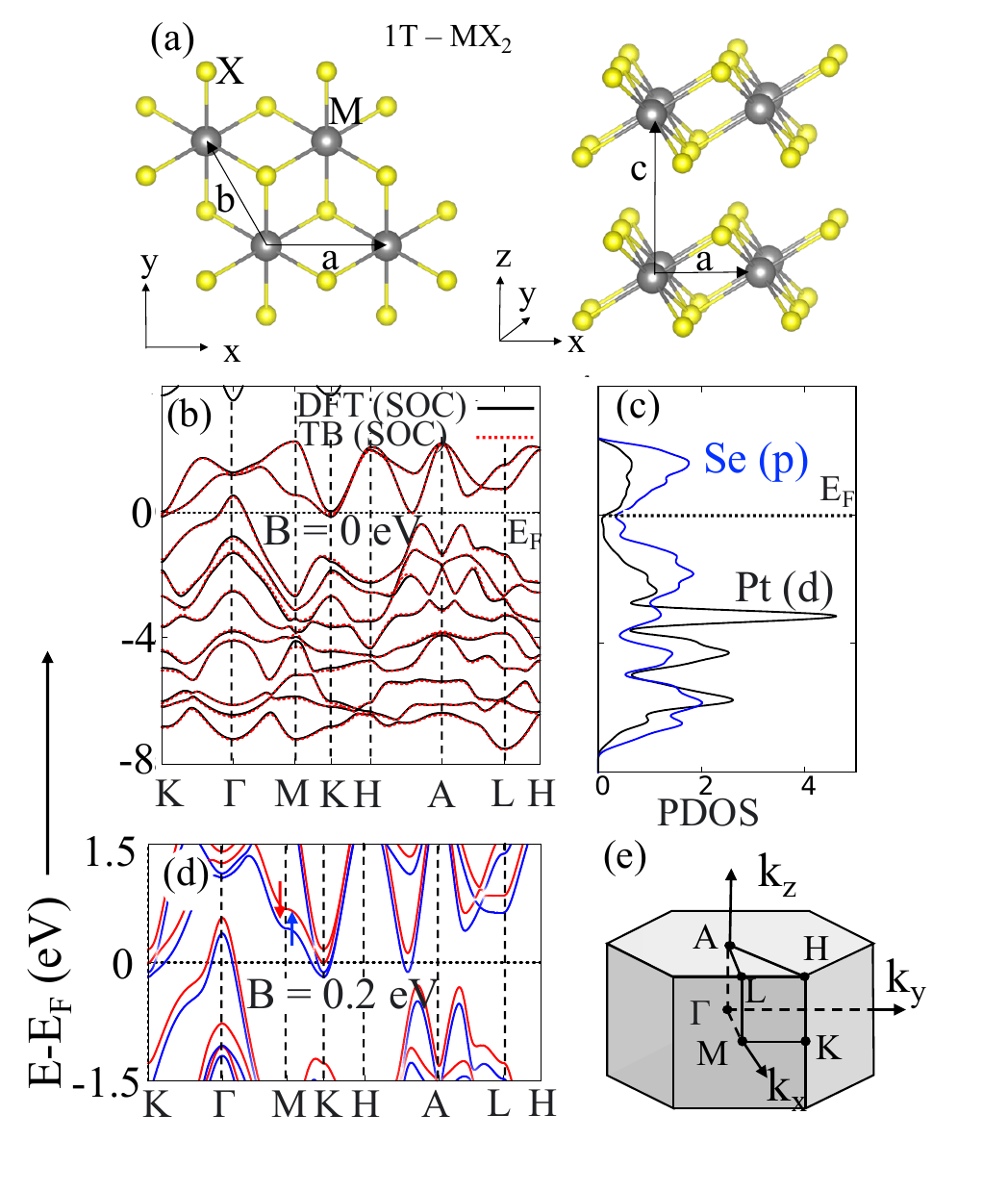}
    \caption{(a) Crystal structure of the 3D TMDs in the centrosymmetric 1T phase, top and side views ({\it left} and {\it right}, respectively).  (b) Comparison of the band structures of  PtSe$_2$ obtained from the DFT and tight binding methods in the presence of SOC and without the magnetic field  ($B=0$). (c) Partial densities of state (PDOS), indicating Pt(d) and Se(p) to be the important orbitals near the Fermi energy. (d) Tight-binding band structure in the presence of both the magnetic field and  SOC.  The colored spin-split bands are not spin pure due to the SOC term,
    which mixes the two spins. Note that the hole pocket at $\Gamma$ and the electron pockets at $K$ as well as along the $H-A$ line are the dominanant contributors to the Hall conductivities due to the energy denominator in the Kubo expressions.
    (e) Sketch of the Brillouin zone. }
    \label{fig:bands}
\end{figure}

We study the 3D TMD compounds in the 1T structure with the formula unit MX$_2$, where M = Pt or  Pd and X = S, Se, or Te. These compounds, with the Space Group No. 164 ($P\bar3 1 m$),  have a trigonal lattice structure [Fig. \ref{fig:bands}(a)] and are centrosymmetric. The band structure near the Fermi energy consists of d orbitals of the M atom and p orbitals of the X atom, and has no band gap. Additionally, there is an electron pocket and a hole pocket at $K$ and $\Gamma$ point, respectively, as shown in Fig. \ref{fig:bands}(b) for PtSe$_2$. 
As we apply an external magnetic field, the spin degeneracy of the bands is broken. As a result, the bands with up and down characters cross the Fermi level at different energies, leading to a spin imbalance. As discussed later, this helps to tune the SHC of the material with the applied magnetic field. Detail results for  materials other than PtSe$_2$ are given in the Supplemental Material\cite{suppl}.

The orbital, spin, and anomalous Hall conductivities (OHC/SHC/AHC) reported here were computed using the DFT using the Quantum Espresso code\cite{QE}, from which a TB model was constructed with the  hopping parameters obtained from the WANNIER90 program\cite{Wannier90}. The spin-orbit coupling (SOC) parameter in the TB Hamiltonian was extracted from the DFT band structure. From the eigenvalues and eigen functions of the TB Hamiltonian, the conductivities were computed using the Kubo formula as discussed later.

We express the TB Hamiltonian  as
\begin{equation}
   {\cal {H}} ={\cal {H}}_0+{\cal {H}}_{SO}+{\cal {H}}_M,
    \label{ham}
\end{equation}
where ${\cal {H}}_0$ is the electron kinetic energy, ${\cal {H}}_{SO}$ is the SOC term, and ${\cal {H}}_{M}$ describes the coupling of the external magnetic field to the  spin and orbital moments of the electron. 
The kinetic energy term is
\begin{eqnarray}
        {\cal H}_0 = \sum_{i m\sigma} \varepsilon_{m}c^{\dagger}_{im\sigma}c_{im\sigma}+\sum_{ijmn\sigma}t_{im,jn}c^{\dagger}_{im\sigma}c_{jn\sigma},
\end{eqnarray}
where $c^\dagger_{i m\sigma}$ is the electron creation operator with $i/ m/ \sigma$ being the site/orbital/spin index. We kept the metal $d^5$ orbitals and the chalcogen $p^3$ orbitals in the basis set, so that we have a $22\times 22$ Hamiltonian including spin, and, furthermore, we kept the hopping integrals beyond nearest neighbors as well. In the Bloch function basis, the matrix elements in the orbital space are spin diagonal and they are given by

\begin{eqnarray}
    {\cal {H}}_{0}^{mn} (\vec k) = \varepsilon_m \delta_{mn} + \sum_{j}t_{mn}^j e^{i\vec{k}\cdot\vec{d_j}},
\end{eqnarray}
where $\varepsilon_m$ is the on-site energy of the $m^{th}$ orbital, and $t_{mn}^j = \braket{0,m|{\cal {H}}|d_j,n}$ denotes the hopping energy between the $m^{th}$ orbital in the central cell and the  $n^{th}$ orbital located at distance $\vec{d_j}$ with respect to  the position of the $m^{th}$ orbital.  
The SOC term is 
\begin{eqnarray}
    {\cal {H}}_{SO} = \frac{2\lambda}{\hbar^2}\vec{L}\cdot\vec{S},
    \label{soc}
\end{eqnarray}
where $\lambda$ is the SOC constant, $\vec{L}$ and $\vec{S}$ are the orbital and spin angular momentum operators. 
For simplicity, we have considered an effective value of $\lambda$ for each material (listed in Table \ref{soc}), so that the TB model reproduces the DFT band structure
in a wide energy region around the Fermi energy. 
The electron spin is  $\vec{S} = (\hbar/2)\vec{\sigma}$, where $\vec{\sigma}$ are the Pauli matrices: $\sigma_x = \begin{pmatrix}
   0&1\\1&0
\end{pmatrix}$, $ \sigma_y = \begin{pmatrix}
   0&-i\\ i & 0
\end{pmatrix}$, and $ \sigma_z = \begin{pmatrix}
   1&0\\ 0 & -1
\end{pmatrix}$. The orbital angular momentum operators for the case of $L = 1$, in the basis set $\phi_p \equiv (p_x,\ p_y,\ p_z)$, are given by
\begin{eqnarray}
    &L_x^{(p)} = \hbar\begin{bmatrix}
        0&0&0\\
        0&0&-i\\
        0&i&0
    \end{bmatrix},
    &L_y^{(p)} = \hbar\begin{bmatrix}
        0&0&i\\
        0&0&0\\
        -i&0&0
    \end{bmatrix},
\\ &L_z^{(p)} =\hbar \begin{bmatrix} \nonumber
        0&i&0\\
        i&0&0\\
        0&0&0
    \end{bmatrix},
    \label{l_p}
\end{eqnarray}
while for  $L=2$, with the d-orbital basis set $\phi_d \equiv (z^2, x^2-y^2, xy, yz, xz)$, they are 
\begin{eqnarray}
    &L_x^{(d)} =\hbar \begin{bmatrix}
        0&0&0&\sqrt{3}i&0\\
        0&0&0&i&0\\
        0&0&0&0&-i\\
        -\sqrt{3}i&-i&0&0&0\\
        0&0&i&0&0
    \end{bmatrix}, \nonumber\\
    &L_y^{(d)} = \hbar\begin{bmatrix}
        0&0&0&0&-\sqrt{3}i\\
        0&0&0&0&i\\
        0&0&0&i&0\\
        0&0&-i&0&0\\
        \sqrt{3}i &-i &0 &0&0
    \end{bmatrix},\\
    &L_z^{(d)} = \hbar\begin{bmatrix}
        0&0&0&0&0\\
        0 &0 &-2i &0 &0\\
        0 &2i &0 &0 &0\\
        0 &0 &0 &0 &i\\
        0 &0 &0 &-i &0
    \end{bmatrix}.\nonumber
    \label{l_d}
\end{eqnarray}

The last term of the Hamiltonian Eq. (\ref{ham}) is  the coupling of the applied magnetic field to the electron magnetic moment, written as
\begin{equation}
    {\cal {H}}_M = \vec{B}\cdot(2\vec{S}+\vec{L}),
    \label{eq8}
\end{equation}
where $\vec B \equiv   \mu_B B_0/\hbar \ \hat z$ is the scaled magnetic field, where the applied magnetic field $B_0$ is taken  along the z direction, and $\mu_B$ is the Bohr magneton. 

The TB band structure obtained from Eq. (\ref{ham}), with parameters extracted from DFT, is compared with the DFT band structure in Fig. \ref{fig:bands}(b), where SOC was present but the magnetic field was zero.
The two band structures agree  very well over the entire Brillouin zone. There is no spin splitting as both the time reversal as well as the inversion symmetries are present. 
With a non-zero magnetic field, the time reversal symmetry is broken, producing the spin splitting in the band structure
(Fig. \ref{fig:bands}(d)). 

\begin{table}[!htb]
    \centering
        \caption{Effective spin-orbit coupling parameter ($\lambda$) for several TMD compounds obtained by fitting with the DFT results. } 
    \begin{tabular}{c |c c c c c c}
\hline\hline

     Material & PtS$_2$    &   PtSe$_2$ & PtTe$_2$ & PdS$_2$ & PdSe$_2$ & PdTe$_2$ \\
\hline
$\lambda\ (eV)$ &0.28 &0.34 &0.38 &0.22 & 0.25& 0.30 \\
\hline\hline
    \end{tabular}
    \label{soc}

\end{table} 

\section{Magnetic moments and anomalous Hall conductivity in presence of a magnetic field}
Due to the presence of both time reversal and inversion symmetries in the TMDs, the momentum space spin/orbital moments as well as the Berry curvature are all zero in the Brillouin zone, but they all become non-zero when the magnetic field is applied, which breaks the time reversal symmetry. As is well known, the non-zero Berry curvature leads to the anomalous Hall effect. In this Section, we study these quantities and evaluate the magnitude of the AHC, to be later compared with the SHC and the OHC.


 \textit{Spin/orbital magnetic moment}: 
 In the semi-classical formulation, the orbital magnetic moment  is expressed as\cite{Niu,Vanderbilt}
\begin{equation}
    \vec{m}^{orb}({\vec{k}}) =\frac{e}{2\hbar c}\sum_n f_{n\vec k}  \bra{\partial_k u_{n\vec{k}}}\times ({\cal {H}}_k+\varepsilon_{n\vec k} - 2\mu)\ket{\partial_k u_{n\vec{k}}},
\end{equation}
where $f_{n\vec k}$ is the Fermi function, $n$ is the band index, $\vec{k}$ is the Bloch momentum, $-e <  0$ is the electron charge,
$\varepsilon_{n\vec k}$ is the band energy, and $u_{n\vec{k}}$ is the corresponding wave function.
 \begin{figure}[!htb]
  \includegraphics[width=0.5\textwidth]{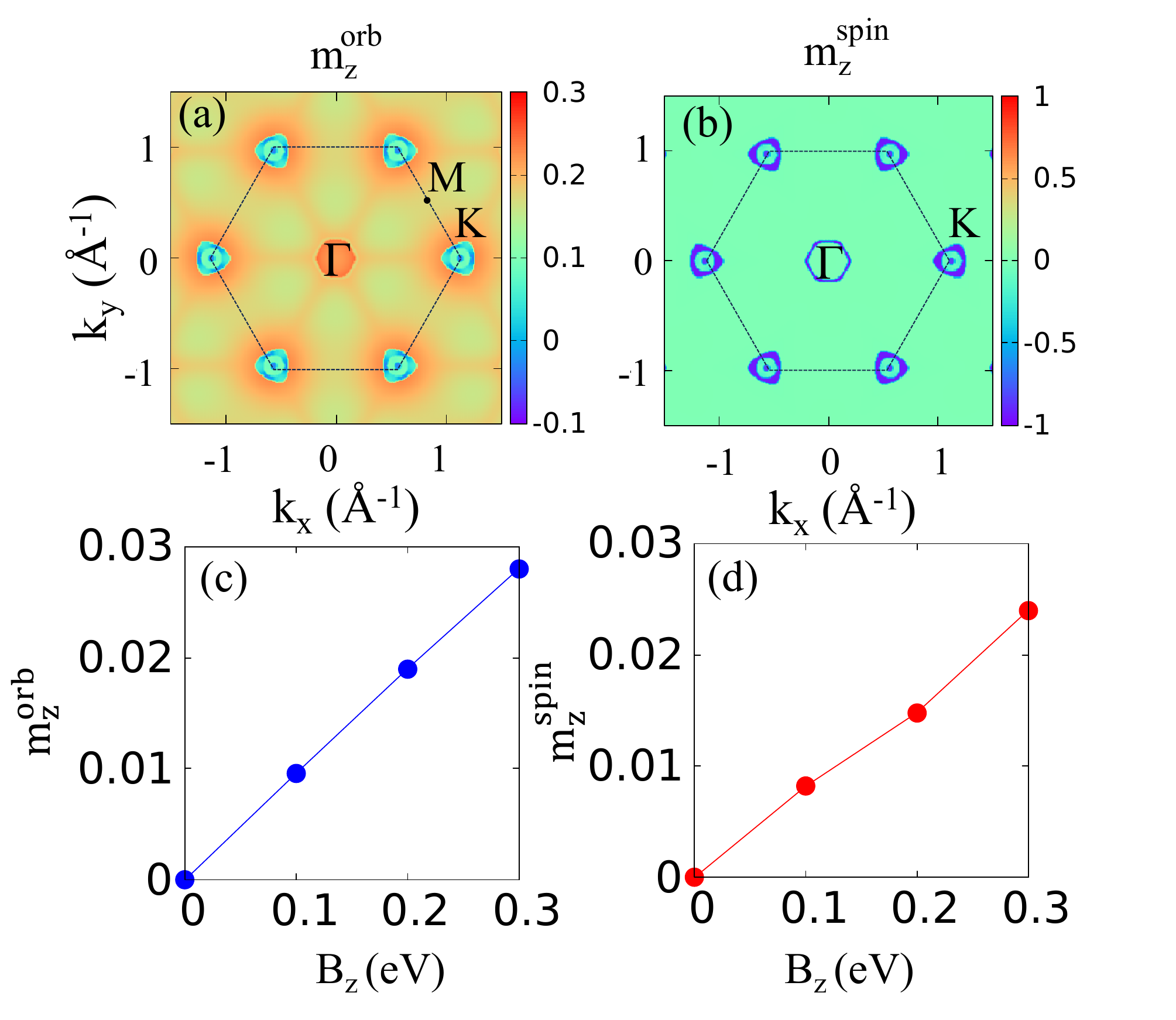}
    \caption{Momentum space distribution of (a)  orbital moment and (b) spin moment for $B_z = 0.1$ eV in  PtSe$_2$ in the $k_z = 0$ plane. The units are in Bohr magneton ($\mu_B$). The moments are prominent around the K and the $\Gamma$ points on this plane, contributing a principal amount to the SHC/OHC. Figs. (c) and (d) show the total orbital and spin moments (BZ sum) as a function of the magnetic field.
    Without a magnetic field, the net moments as well as the moments everywhere in the BZ are zero due to the existence of both the inversion and the time reversal symmetry.
    }
    \label{moments}
\end{figure}
It is known that in presence of the inversion symmetry, the orbital magnetic moment follows $m^{orb}(\vec{k}) = m^{orb}(-\vec{k})$, whereas when the time reversal symmetry is present, one gets $m^{orb}(\vec{k}) =- m^{orb}(-\vec{k})$. Hence, the presence of both the time reversal and inversion symmetries nullifies the magnetic moment throughout the BZ. However, by breaking one of the two symmetries (in our case breaking the time reversal symmetry with the magnetic field), a non-zero distribution of the moments can be obtained in the momentum space. The same is true for both the spin moment as well as for the Berry curvature.
We compute the spin moment from the expectation value of the spin operator $\vec{S}$
\begin{equation}
    \vec{m}^{spin}({\vec k}) = -g_s\frac{e}{2m_e}\sum _{n} f_{n\vec k}\braket{u_{n\vec k}|\vec{S}|u_{n\vec k }}.
\end{equation}

The computed values of the spin/orbital moments for PtSe$_2$ in the presence of a magnetic field are shown in
Fig. \ref{moments}. As expected, they are zero everywhere in the BZ in the absence of a magnetic field. 
Note that the spin moments disappear if the two spin-split bands are both occupied.
Thus as may be inferred from the band structure of Fig. \ref{fig:bands}(b), there is spin moment only near the $\Gamma$ and $K$ points in the BZ, as seen from Fig. \ref{moments} (b).



\textit{Anomalous Hall conductivity:}
Before we study the SHC and OHC, we compute the magnitude of the AHC,
which can be expressed as the sum of the Berry curvatures over the occupied band states in the Brillouin zone\cite{Niu}, viz.,
\begin{equation}
\sigma^{\rm AHE}_{\alpha \beta}   =  -\frac{e^2} {\hbar N_k V_c} \sum_{n \vec  k}^{occ} \Omega_{n\vec  k}^\gamma, 
\end{equation} 
where $(\alpha, \beta, \gamma)$ are cyclic permutations of $(x,y,z)$, $V_c$ is the unit cell volume, and $N_k$ is the number of $k$-points in the BZ.  As usual, the conductivity is defined from the expression for the current density 
$j_{\alpha\beta} = \sigma_{\alpha \beta} E_\beta$, where the current density $j_{\alpha\beta}$ is along the $\alpha$ direction, generated
by the electric field along the $\beta$ direction.
The Berry curvature may be expressed using the Kubo formula
\begin{eqnarray}
\Omega_{n\vec  k}^\gamma = -2\hbar^2\ Im\sum_{m\neq n}\frac{\braket{u_{n\vec{k}}|v_\alpha|u_{m\vec{k}}}\braket{u_{m\vec{k}}|v_\beta|u_{n\vec{k}}}}{(\varepsilon_{m\vec{k}}-\varepsilon_{n\vec{k}})^2},
\end{eqnarray}
where    $v_\alpha = \hbar^{-1}(\partial {\cal {H}}/\partial k_\alpha)$ is the velocity operator. 
Note that the Berry curvature also follows  similar symmetry arguments as for the spin/orbital magnetic moment, already discussed above. 
It therefore vanishes in the absence of the magnetic field, since the system has both time reversal and inversion symmetries. However, an applied magnetic field will break the time reversal symmetry and create a non-zero Berry curvature. 
This leads to a non-zero anomalous velocity given by $e \hbar^{-1}\vec{E}\times \vec \Omega_{n\vec{k}}$, causing a transverse flow of electrons, resulting in the anomalous Hall effect (AHE). 
Since the charges also carry intrinsic orbital and spin magnetic moments, this also leads to a orbital/spin accumulation across the material, leading to the OHE/SHE, which we will study in the next Section. 

\begin{figure}[!htb]
    \centering
    \includegraphics[width=0.5\textwidth]{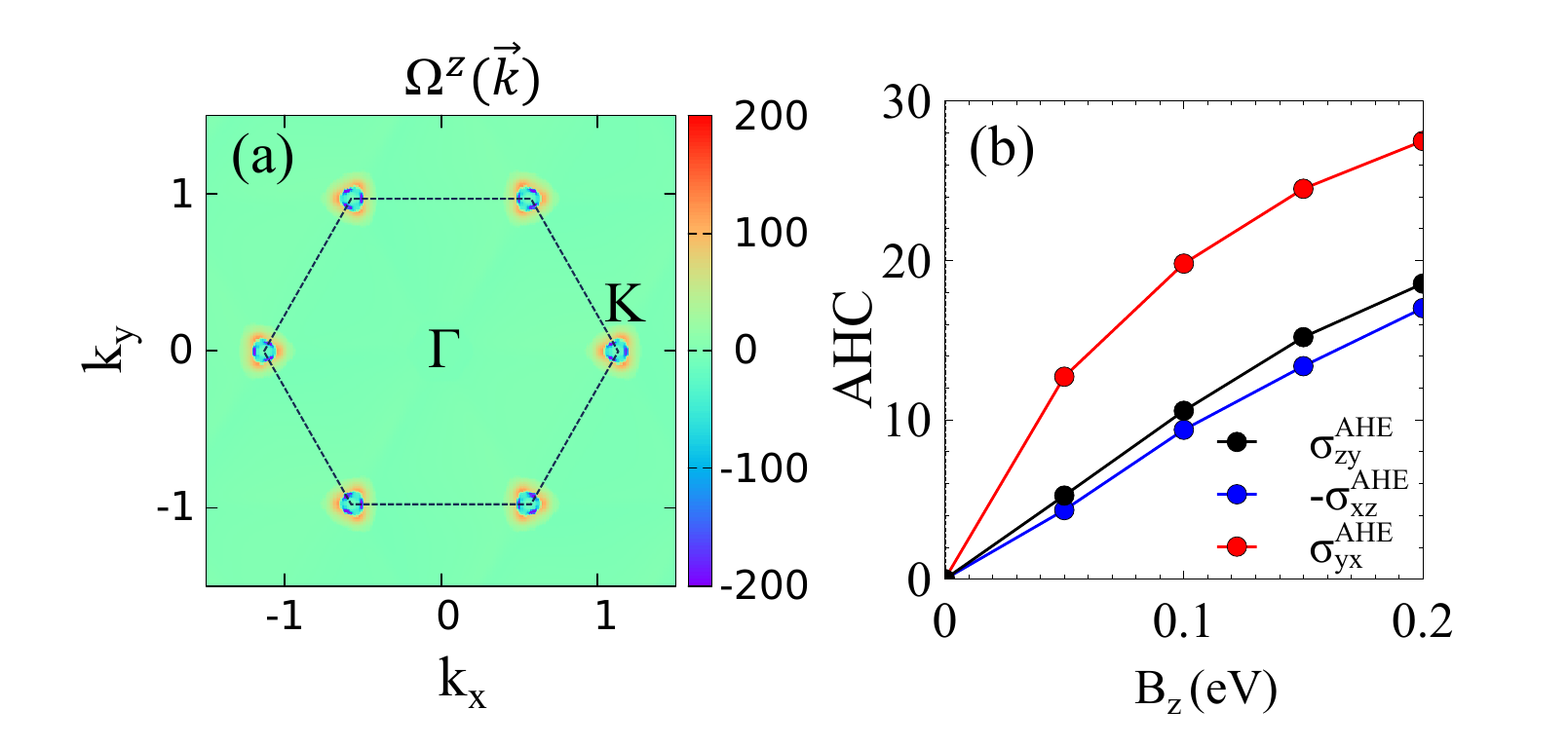}
    \caption{ Anomalous Hall conductivity in PtSe$_2$ in the presence of the magnetic field. (a) Berry curvature $\Omega^z (\vec k)$ (units of \AA$^2$), defined as the sum of $\sum_n \Omega^z_{n \vec k}$ over the occupied bands $n$, on the $k_z=0$ plane with the scaled magnetic field 
    $\mu_B B_0/\hbar \equiv B_z = 0.1$ eV. (b) Different components of the AHC  in  units of $\hbar/e\ \Omega^{-1}cm^{-1}$ as a function of the magnetic field. 
}
    \label{ber_bz}
\end{figure}

The computed Berry curvature and the AHC are shown in Fig. \ref{ber_bz}.
From the figure, we see that large values of the Berry curvature occur at the momentum points, where bands cross 
the Fermi energy (Fig. \ref{fig:bands} (d)) due to the energy denominator in the Kubo formula. The AHC is non-zero only when a magnetic field is present, as expected.

\section{Orbital and spin Hall conductivities}

The key quantities we study in this paper are the  orbital and spin Hall conductivities in the presence of a magnetic field, 
which are important ingredients in the generation of the spin-orbit torque in the TMD heterostructures with magnetic materials, which has been studied in several experiments\cite{pranaba,she-exp-ptte2}. 

We compute the OHC and SHC from the momentum sum of the orbital/spin Berry curvatures (OBC/SBC)
\begin{equation}\label{OHC} 
  \sigma^{\gamma,\rm orb/spin}_{\alpha \beta}   =  -\frac{e} { N_k V_c} \sum_{ \vec  k} \Omega^{\gamma,\rm orb/spin}_{\alpha \beta} ({\vec  k}),
\end{equation}
where $\Omega^{\gamma,\rm orb/spin}_{\alpha \beta} ({\vec  k})
= \sum_n^{\rm occ} \Omega^{\gamma,\rm orb/spin}_{n,\alpha \beta} ({\vec  k})$ is the sum over occupied bands $n$ at a specific momemtum point $\vec k$.
The OBC for the Bloch state $n\vec k$ is computed using the Kubo formula
\begin{equation} \label{obc}         
 \Omega^{\gamma,\rm orb}_{n,\alpha\beta} ({\vec  k}) = 2 \hbar   \sum_{n^\prime \neq n} \frac {{\rm Im}[ \langle u_{n{\vec  k}} | \mathcal{J}^{\gamma,\rm orb}_\alpha | u_{n^\prime{\vec  k}} \rangle  
\langle u_{n^\prime{\vec  k}} | v_\beta | u_{n{\vec  k}} \rangle]} 
{(\varepsilon_{n^\prime \vec k}-\varepsilon_{n  \vec k} )^2},
\end{equation}
where $v_{\alpha} =  \frac{1}{\hbar} \frac{\partial {\cal H} }{ \partial k_\alpha}$ is the velocity operator, and the orbital current operator is $\mathcal{J}^{\gamma,\rm orb}_\alpha = \frac{1}{2} \{v_\alpha, L_\gamma \}$,  
with $L_\gamma$ being the orbital angular momentum. For the SBC, the orbital current operator is replaced by the spin current operator $\mathcal{J}^{\gamma,\rm spin}_\alpha = \frac{1}{2} \{v_\alpha, S_\gamma \}$ in Eq. (\ref{obc}). Sufficient number of $k$ points are taken in the momentum sum in Eq. (\ref{OHC}) to ensure convergence, which typically required 
a $k$-mesh of $120\times120\times120$ points in the full BZ. 

  %
  \begin{figure}[!htb]
    \includegraphics[scale = 0.4]{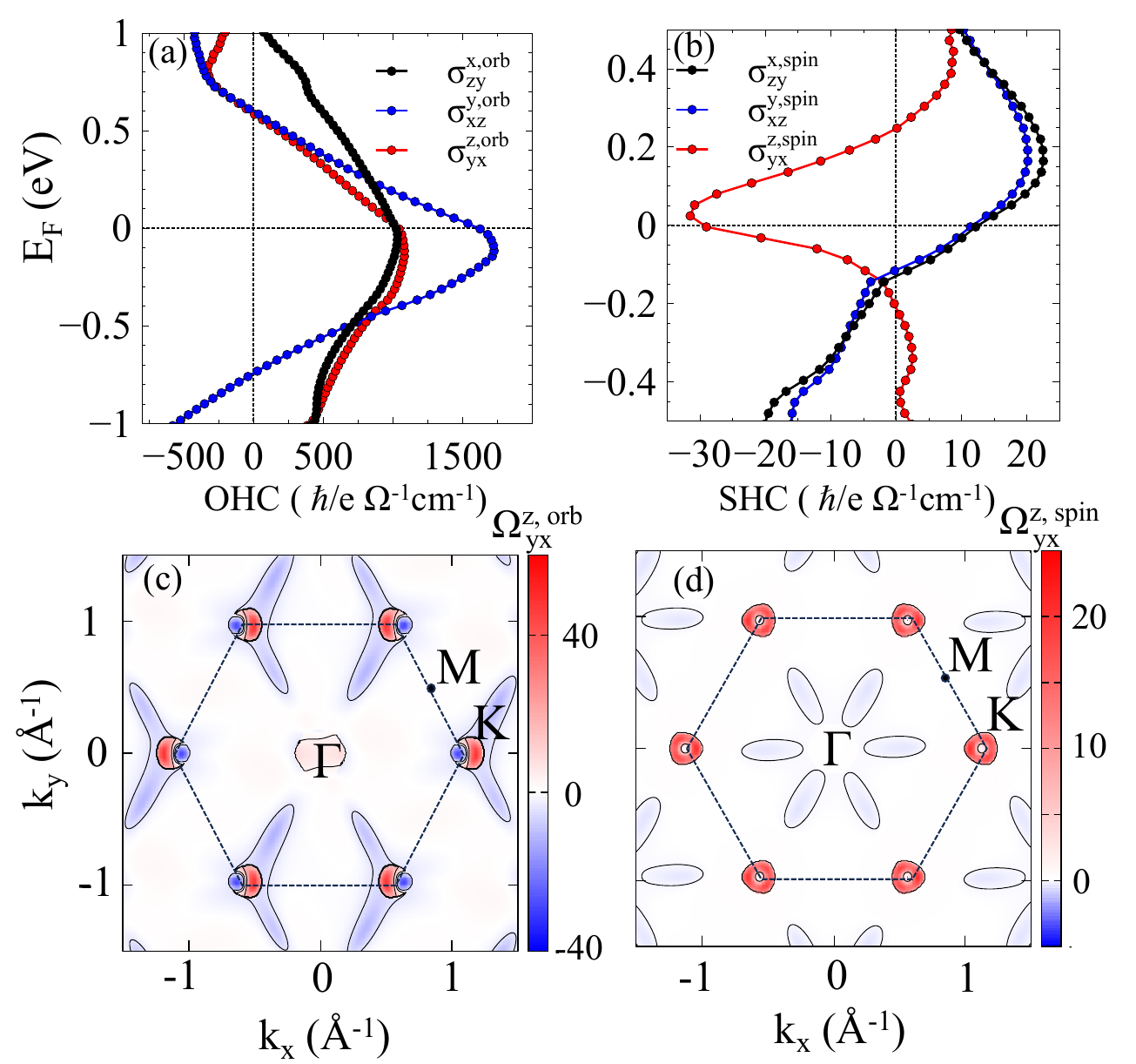}
    \caption{(a) Orbital and (b) spin Hall conductivity for  PtSe$_2$, with $B = 0$, as a function of the Fermi energy. (c) and (d),  orbital and spin Berry curvature sums $\Omega^{z,\rm orb/spin}_{yx} (\vec k)$,
    defined as the sum over occupied bands of the corresponding Berry curvature at a specific $\vec k$ point, in units of \AA$^2$ in the $k_z = 0$ plane for the case $E_F = 0$. Contours of constant OBC/SBC values are shown in Figs. (c) and (d). 
    }
    \label{ohc_shc_ptse2}
\end{figure}
\begin{figure}[!htb]
    \centering
    \includegraphics[width = 0.35\textwidth]{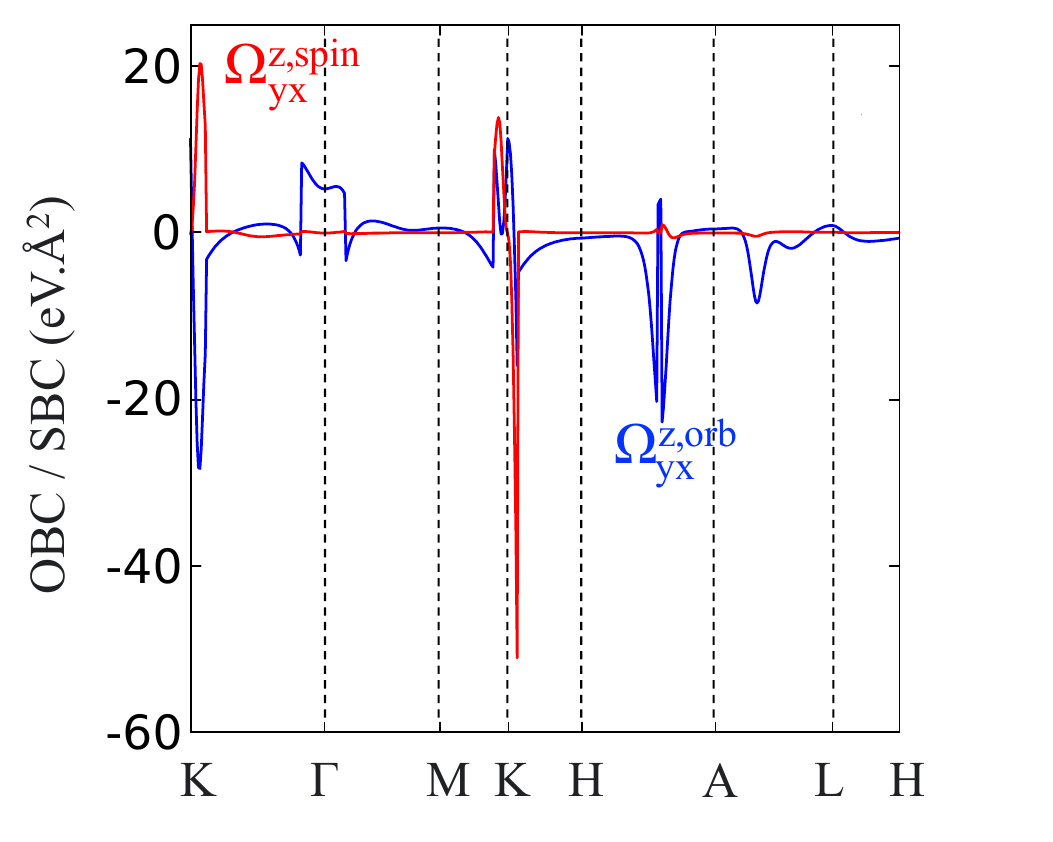}
    \caption{Orbital and spin Berry curvature sums along the high symmetry lines for PtSe$_2$. 
    }
    \label{obcline}
\end{figure}

\begin{figure}[!htb]
    \centering
    \includegraphics[width = 0.38\textwidth]{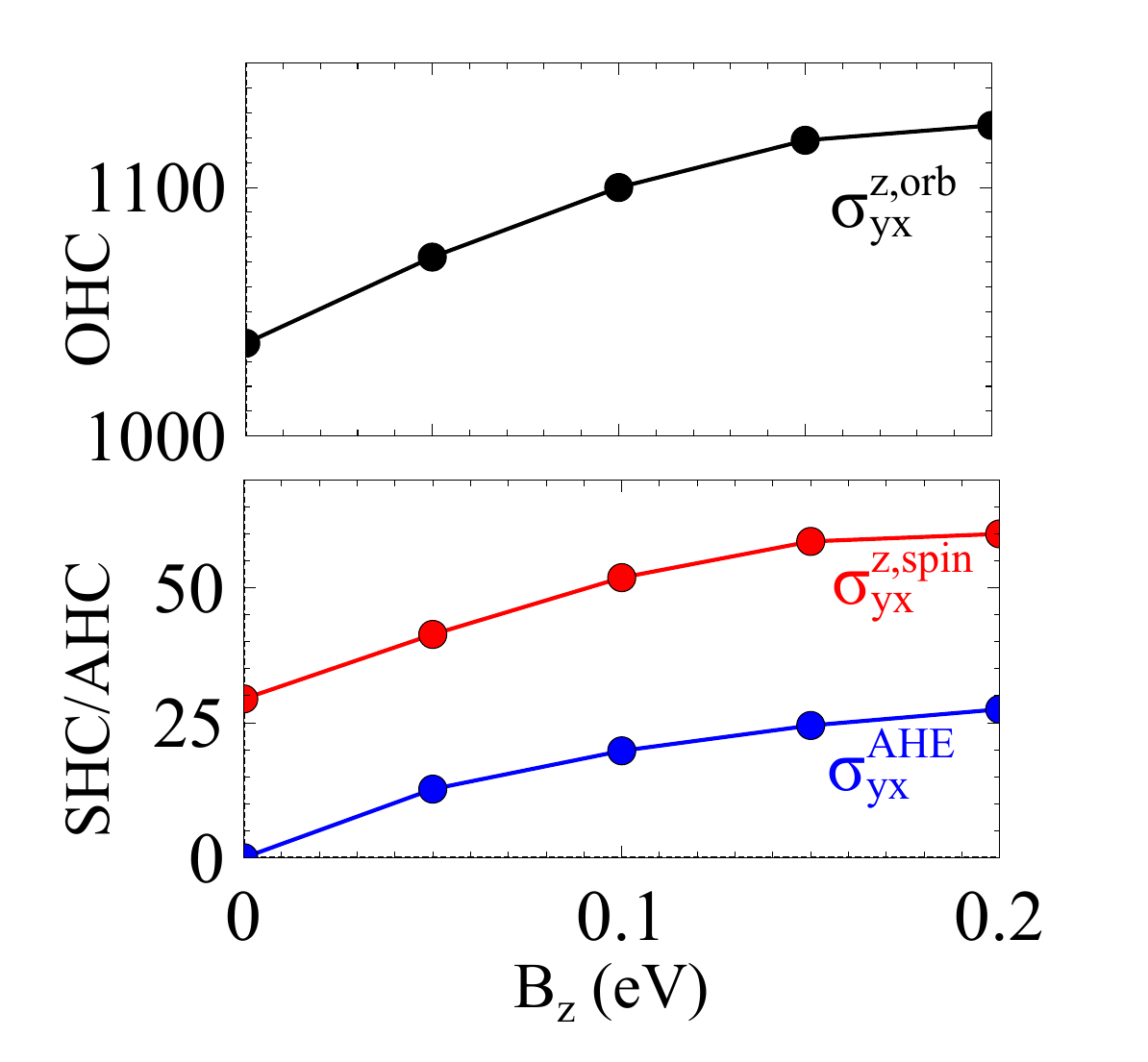}
    \caption{Orbital, spin, and anomalous Hall conductivities for PtSe$_2$ as a function of the magnetic field.
    Note that the magnitude of the OHC is by far the largest as compared to the other two, and of course the AHC is non-zero only when a magnetic field is present. 
    The conductivities are in  units of $\hbar/e\ \Omega^{-1} cm^{-1}$. 
    }
    \label{cond_bz_all}
\end{figure}

Fig. \ref{ohc_shc_ptse2} shows the computed OHC and SHC for PtSe$_2$ as a function of the Fermi energy.
Once again, the major contribution to the conductivity comes from the $\Gamma$ and $K$ valleys of the BZ as indicated from the Berry curvatures  shown in Figs. \ref{ohc_shc_ptse2}(c) and (d), as well as
from Fig. \ref{obcline}, which shows the Berry curvatures along the symmetry lines in the BZ. The last figure also shows additional substantial contributions from momentum points along the  H-A line. 
One point to note from Fig. \ref{ohc_shc_ptse2} is that the SHC is an order of magnitude less than the OHC.
In fact, the OHC dominates the SHC even when a magnetic field is present as summarized in Fig. \ref{cond_bz_all}.

However, in spite of its smaller magnitude, the spin current is thought to play the dominant role in magnetic torque experiments in typical non-magnet/FM bilayers. For example, in the torque experiments on FM/Pt bilayers\cite{ot-3}, it has been estimated that the orbital-to-spin conversion efficiency is typically only a few percent (1-5 \%), so that overall, the orbital current is not as effective as the spin current in producing the torque, though it is much stronger in magnitude.

\begin{figure}[!htb]
    \centering
    \includegraphics[width = 0.5\textwidth]{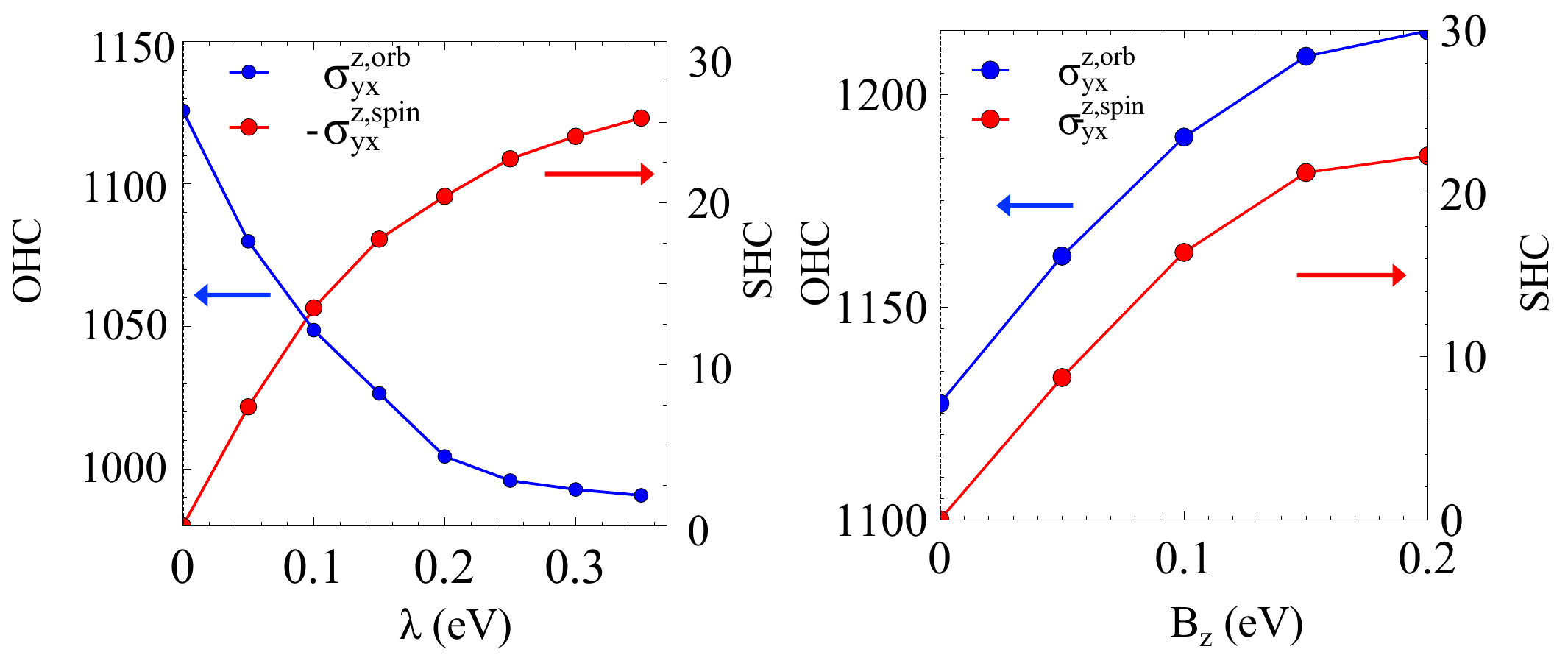}
        \caption{ (a) Dependence of the OHC and SHC for  PtSe$_2$ (units of $\hbar/e\ \Omega^{-1}cm^{-1}$) on the SOC strength $\lambda$. Here, $B_z = 0$. (b) The same as a function of the magnetic field but with  $\lambda = 0$.
        The SHC is zero, if both $\lambda$ and $B_z$ are zero. 
        }
    \label{cond_la}
\end{figure}

We note in passing that the OHE is considered more
fundamental than the SHE in the sense that it can exist without the SOC $\lambda$.
The OHE can originate from the orbital texture, created in response to the electric field\cite{Go}, and neither the SOC nor a magnetic field is necessary for its existence. In contrast, one of these two needs to be present for the SHE, simply because otherwise there is no coupling between the spin space and the real space. This is clearly seen from Fig. \ref{cond_la}, where we have shown the dependence of the OHC and the SHC on the 
SOC strength $\lambda$ and the magnetic field.
The SHC is zero unless there is a non-zero $\lambda$ or a magnetic field.

 The Hall conductivities for several TMD compounds without any magnetic field are listed in Table \ref{cond}. In all cases, we find that the OHC is much higher than the SHC by many orders of magnitude.


\begin{table}[!htb]
    \centering
        \caption{Orbital and spin Hall conductivities (in units of $\hbar/e\ \Omega^{-1} cm^{-1}$) for  several  TMD compounds with $B = 0$. Note that OHC $ >> $ SHC for all cases.} 
    \begin{tabular}{c| c c c| c c c}
\hline\hline
&\multicolumn{3}{c|}{OHC}&
\multicolumn{3}{c}{SHC}\\
\hline
     Material & $\sigma_{zy}^{x,orb}$    &  $\sigma_{xz}^{y,orb}$& $\sigma_{yx}^{z,orb}$ &$\sigma_{zy}^{x,spin}$ &$\sigma_{xz}^{y,spin}$ & $\sigma_{yx}^{z,spin}$ \\
\hline
PtS$_2$ &799 & 1545 & 1049 &0.07  &-0.91  &-1.12  \\

PtSe$_2$ & 1021& 1601 & 1037 &12.8  &11.9  & -29.4 \\
PtTe$_2$ & 602& 819 & 1168 &-38.2  &  -27.07&-50.7  \\
PdS$_2$ &858 &-146  & -880& -0.01 & 0.49 & 0.05 \\
PdSe$_2$ & 840& -126  & -803 & -5.48 & 20.05 & 7.42 \\
PdTe$_2$ &824 &-112  &-784  & -35.4 &-22.1  &15.04  \\
\hline\hline
    \end{tabular}
\label{cond}

\end{table}

\textit{Effect of a magnetic field} --
Since in the magnetic torque experiments on the non-magnetic/FM bilayers, a strong magnetic field exists in the non-magnetic material due to the proximity effect, it is important to consider how the magnetic field might affect the Hall conductivities.  A main result of this Section is that we find the SHC to increase by as much as a factor of two  for typical fields that are present near the interface, while the OHC increases only by a modest amount. The substantial effect of the proximity magnetic field on the spin current must therefore be taken into account in interpreting the experiments, since the spin current is expected to be important in the generation of the magnetic torque. 

In our calculations, we applied a magnetic field in the $z$ direction, i.e., along the c axis of the material. The magnetic field has two effects that enhance the Hall conductivities. First, it creates a net orbital/spin moment in the sample,
with momentum dependent values over the BZ, as indicated in Fig. \ref{moments}. Second, a strong spin polarization is produced near the valley points K, and to some extent also at the $\Gamma$ point, where the bands cross the Fermi energy, due to the Zeeman spin splitting produced by the magnetic field. It turns out that this spin imbalance has a strong effect  on the SHC, which is enhanced by a factor of two or so for magnetic fields that exist near the interface with a FM.
The OHC is also enhanced, but by a much smaller amount. 

To be more specific, in these systems, the band structure, shown in Fig. \ref{fig:bands}, has an electron pocket at the  $K$ point and a hole pocket at $\Gamma$. As the magnetic field is applied, it induces spin splitting of the band structure, creating an imbalance between up and down spins near these high symmetry points. Since the SBC for the bands with spin up character is almost opposite to that of the spin down bands for a specific k point, the imbalance created by the magnetic field leads to a net increase of the SBC and in turn of the SHC.
\begin{figure}[!htb]
    \centering
    \includegraphics[width = 0.5\textwidth]{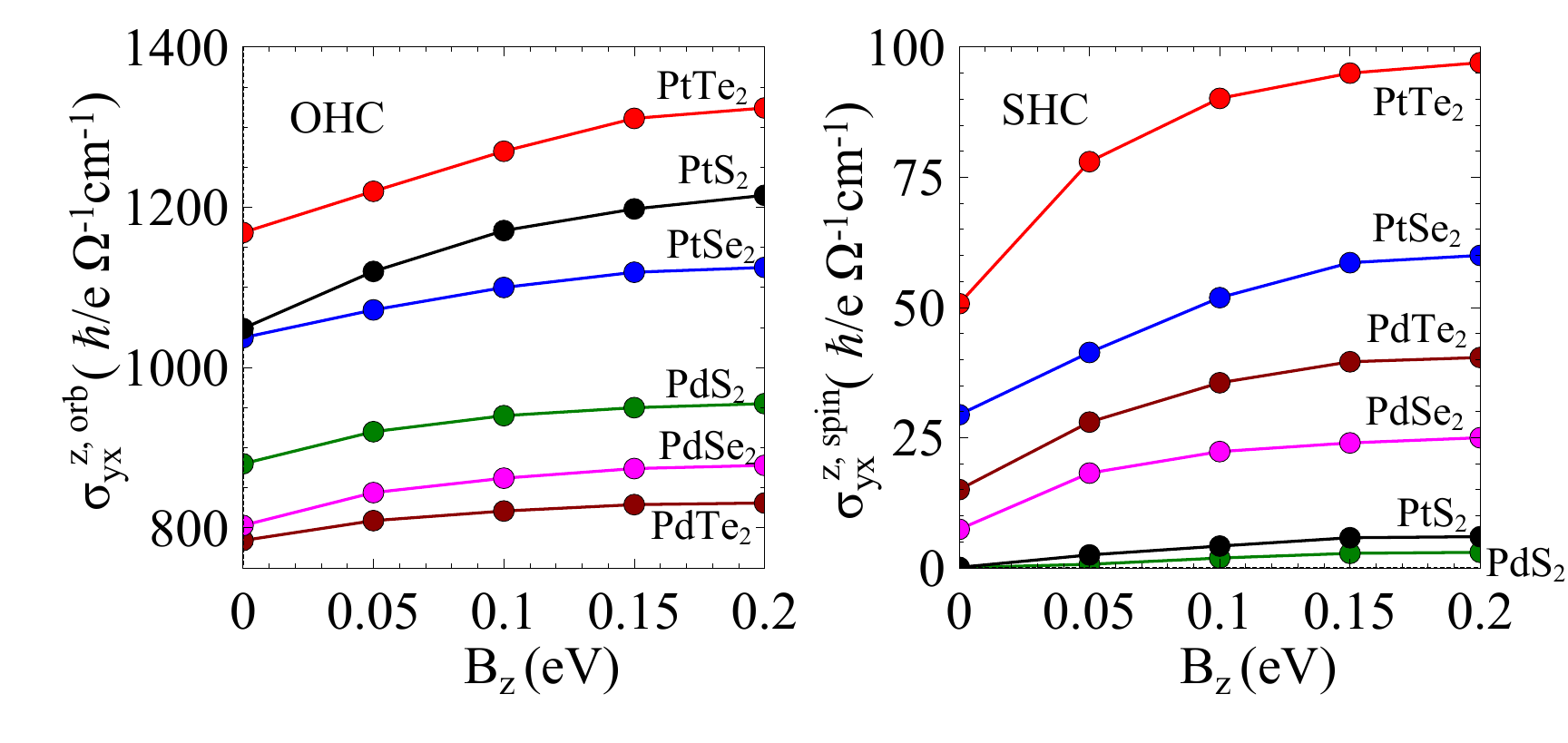}
    \caption{ Magnitudes of the OHC (left) and the SHC (right) as a function of the applied magnetic field for several TMD compounds.}
    \label{cond_bz}
\end{figure}

The computed values of the  OHC/SHC as a function of the magnetic field for several TMD compounds  are shown in Fig. \ref{cond_bz}.
As we have already mentioned, the OHC is significantly larger than the SHC for all compounds and for all magnetic fields.
A second point to note is that one might naively think that the larger the SOC, the larger would be the SHC. This would be generally true if the SOC strength $\lambda$ can be increased within a certain material as was shown in Fig. \ref{cond_la} (a) for PtSe$_2$, however this is not the case while comparing between different materials.
This is because the details of the band structure,
in particular, the band crossings at the Fermi energy, have a dominant effect on the SHC. 
As a result, though the effective $\lambda$ as seen from Table \ref{soc} are not that different, the magnitudes of the SHC are different, as seen from Table II and Fig. \ref{cond_bz}, by an order of magnitude between the materials considered.

\section{Summary and outlook}
In summary, we studied the orbital and spin Hall effects in a series of bulk 1T transition-metal dichalcogenides (TMDs) of the form MX$_2$:  M = Pt, Pd and X = S, Se, and Te, in the presence of a magnetic field. Our work was motivated by the fact that many spintronics experiments involve non-magnetic/ferromagnetic heterostructures, where the spin Hall effect of the non-magnetic material is used to inject spin current into the ferromagnet. We used the TMDs as a prototype because these have been proposed to be effective non-magnetic materials in such devices.

In all materials we studied, the orbital Hall conductivity (OHC) was found to be two to three orders of magnitude larger than the SHC. Concering the SHC, it is large in PtTe$_2$ and PdTe$_2$ as compared to the other compounds suggesting the usefulness of these materials for spintronics devices. In magnetic torque experiments involving
the non-magnetic/ferromagnetic heterostructures so far studied, due to weak effective spin-orbit correlation in the ferromagnetic domain, the orbital current has been found rather inefficient in producing the torques, even though it may be much larger in magnitude. If this issue can be resolved, this may make the TMDs with large OHC to be important materials for a variety of spintronics devices.


In the presence of a magnetic field, such as the 
proximity field of the ferromagnet in the heterostructure, we found the SHC to be enhanced by quite a large amount, as much as a factor of two, while the OHC increased by a modest amount. Thus in device applications, the effect of any magnetic 
field present must be taken into account in 
determining the magnitude of the SHC/OHC.
Additionally, in free standing samples, the OHC/SHC may be tuned by the application of an external magnetic field.
Our work would be important in the interpretation  of the spin-charge conversion experiments in spintronic devices such as  SOT-MRAM, spin nano-oscillators, spin logic devices etc., involving the TMD materials.

\section*{ACKNOWLEDGEMENT}
This research was supported by Science and
Engineering Research Board of Department of Science
and Technology (SERB-DST), India through Grant No.
CRG/2020/004330. Pratik Sahu thanks IIT Madras for the financial support through institute of eminence (IoE) scheme. The authors thank Pranaba Muduli for useful discussions.




\begin{thebibliography}{10}

\bibitem{she-2} T. Tanaka, H. Kontani, M. Naito, T. Naito, D. S. Hirashima, K. Yamada and J. Inoue, Intrinsic spin Hall effect and orbital Hall effect in 4d and 5d transition metals, Phys. Rev. B \textbf{77}, 165117 (2008).

\bibitem{she-3} H. Kontani, T. Tanaka, D. S. Hirashima, K. Yamada and J. Inoue, Giant Intrinsic Spin and Orbital Hall Effects in Sr2MO4 (M = Ru, Rh, Mo), Phys. Rev. Lett. \textbf{100}, 096601 (2008).

\bibitem{she-1}H. Kontani, T. Tanaka, D. S. Hirashima, K. Yamada and J. Inoue, Giant Orbital Hall Effect in Transition Metals:
Origin of Large Spin and Anomalous Hall Effects, Phys. Rev. Lett. \textbf{102}, 016601 (2009).



\bibitem{Go} D. Go, D. Jo, C. Kim, and H-W. Lee, Intrinsic Spin and Orbital Hall Effects from Orbital Texture, Phys. Rev. Lett. {\bf 121}, 086602 (2018).

\bibitem{Jo} D. Jo, D. Go and H.-W. Lee, Gigantic intrinsic orbital Hall effects in weakly spin-orbit coupled metals, Phys. Rev. B {\bf 98}, 214405 (2018).

\bibitem{sayantika_1} S. Bhowal and S. Satpathy, Intrinsic orbital moment and prediction of a large orbital Hall effect
in two-dimensional transition metal dichalcogenides, 
Phys. Rev. B {\bf 101}, 121112(R) (2020).

\bibitem{sayantika_2} S. Bhowal and S. Satpathy, Intrinsic orbital and spin Hall effects in monolayer transition metal dichalcogenides, Phys. Rev. B {\bf 102}, 035409 (2020). 

\bibitem{pratik-ohe} P. Sahu, S. Bhowal, and S. Satpathy, Effect of the inversion symmetry breaking on the orbital Hall effect: A model study, Phys. Rev. B \textbf{103}, 085113 (2021).

\bibitem{she-exp} L. Wang, R. J. H. Wesselink, Y. Liu, Z. Yuan, K. Xia, and P. J. Kelly, Giant room temperature interface spin Hall and inverse spin Hall effects, Phys. Rev. Lett. {\bf 116}, 196602 (2016).

 \bibitem{ot-1} D. Go and H. W. Lee, Orbital torque: Torque generation by orbital current injection, Phys. Rev. Research {\bf 2}, 013177 (2020).


\bibitem{ot-2} Z. C. Zheng, Q. X. Guo, D. Jo, D. Go, L. H. Wang, H. C. Chen, W. Yin, X. M. Wang, G. H. Yu, W. He, H. W. Lee, J. Teng, and T. Zhu, Magnetization switching driven by current-induced torque from weakly spin-orbit coupled Zr, Phys. Rev. Research {\bf 2}, 013127 (2020).

\bibitem{ot-3} D. Lee, D. Go, H-J. Park, W. Jeong, H-W. Ko, D. Yun,
D. Jo, S. Lee, G. Go, and J. H. Oh et al., Orbital torque in magnetic bilayers, Nat Comm. {\bf 12}, 6710 (2021).

\bibitem{rap1} L. M. Canonico, T. P. Cysne, T. G. Rappoport, and R. B. Muniz, Two-dimensional orbital Hall insulators, Phys. Rev. B {\bf 101}, 075429 (2020).

\bibitem{rap2} L. M. Canonico, T. P. Cysne, A. Molina-Sanchez, R. B. Muniz,
and T. G. Rappoport, Orbital Hall insulating phase in transition
metal dichalcogenide monolayers, Phys. Rev. B {\bf 101}, 161409(R)
(2020).

\bibitem{pranaba} R. Mudgal \textit{et al}., Magnetic-Proximity-Induced Efficient Charge-to-Spin Conversion in Large-Area PtSe2/Ni$_{80}$Fe$_{20}$ Heterostructures, Nano Lett. 2023, \textbf{23}, 24 (2023).

\bibitem{OHE-3-Cr} I. Lyalin, S. Alikhah, M. Berritta, P. M. Oppeneer, and R. K. Kawakami, Magneto-Optical Detection of the Orbital Hall Effect in Chromium, Phys. Rev. Letts. {\bf 131}, 156702 (2023).

\bibitem{OHE-4-Ti} Y. G. Choi, D. Jo, K. H. Ko, D. Go, K. H. Kim, H. G. Park, C. Kim, B. C. Min, G. M. Choi, and H. W. Lee, Observation of the orbital Hall effect in a light metal Ti, Nature {\bf 619}, 52 (2023).




\bibitem{suppl} See supplemental material for the extra results and figures.

\bibitem{QE} P. Giannozzi, S. Baroni, N. Bonini, M. Calandra, and R. Car, et. al., QUANTUM ESPRESSO: a modular and open-source software project for quantum simulations of materials, J. Phys. Condens. Matter {\bf 21}, 395502 (2009).
\bibitem{Wannier90} G. Pizzi, V. Vitale, R. Arita, S. Blügel, and F. Freimuth, et. al., Wannier90 as a community code: new features and applications, J. Phys. Condens. Matter {\bf 32}, 165902 (2020).

\bibitem{Niu} D. Xiao, J. Shi and Q. Niu, Berry Phase Correction to Electron Density of States in Solids, Phys. Rev. Lett. {\bf 95}, 137204 (2005); D. Xiao, M.-C. Chang, and Q. Niu, Berry phase effects on electronic properties, Rev. Mod. Phys. {\bf 82}, 1959 (2010). 

\bibitem{Vanderbilt} D. Ceresoli, T. Thonhauser, D. Vanderbilt, and R. Resta, Orbital magnetization in crystalline solids: Multi-band insulators, Chern insulators, and metals,  Phys. Rev. B {\bf 74}, 024408 (2006).

\bibitem{she-exp-ptte2} H. J. Xu, J. W. Wei, H. A. Zhou, J. F. Feng, T. Xu, H. F. Du, C. L. He, Y. Huang, J. W. Zhang, and Y. Z. Liu, et al., High Spin Hall Conductivity in Large-Area Type-II Dirac Semimetal PtTe$_2$. Adv. Mater. 2020, {\bf 32}, 2000513 (2020).






















\end{thebibliography}
\end{document}